\documentstyle[twocolumn,aps,epsfig]{revtex}
\begin{document}
\draft
\title{Hadronic Equipartition of Quark and Glue Momenta}
\author{Y.N. Srivastava and A. Widom}
\address{Physics Department, Northeastern University, Boston MA 02115}
\maketitle

\begin{abstract}
If the ``glue'' which binds quarks within hadrons takes the form 
of strings, then a virial theorem may be derived which shows 
how the total hadron four momentum splits up into a quark contribution 
plus a glue contribution. The hadrons made up of light quarks exhibit 
an equipartion of four momentum into equal parts quarks and glue. 
The agreement with the experimental ``parton'' distribution four 
momentum sum rule is quite satisfactory as is the string fragmentation 
model.
\end{abstract}  

\narrowtext

\section{Introduction} 

The quark model presently plays a role in the  classification 
of hadronic particles similar to the role played by the periodic 
table of chemical elements in the nineteenth century. From a 
microscopic viewpoint, the interaction holding the quarks 
together is often taken to be described by the Yang-Mills 
glue of quantum chromo-dynamics (QCD). However, the QCD derivation 
of the quark confinement within the hadron is more than just a 
little obscure. In order to make contact with experiments, some 
effective model of confinement must be constructed. The models 
include bags or string fragmentation constructions. The word 
``glue'', within a given model, here denotes whatever the physical 
substance (perhaps Yang-Mills gauge gluons) that binds the quarks 
together.

Of the QCD inspired quark binding pictures, perhaps the most popular 
among experimentalists is the string fragmentation model
\cite{1} \cite{2}. This 
model can be partly understood as a consequence of the Wilson closed 
loop conjecture\cite{3}. Briefly, Wilson considered the action 
of sending a quark 
with QCD charge \begin{math} g \end{math} around a closed 
space-time path \begin{math} \partial \Sigma \end{math} which bounds 
a space-time area \begin{math} \Sigma \end{math}; i.e the 
vacuum Wilson action \begin{math} W \end{math} is defined by 
\begin{equation}
\exp\left({iW\over \hbar }\right)=\left<0\left|\ \exp
\left({ig\over \hbar c}
\int_{\partial \Sigma }(A^c_\mu \lambda_c/2)dx^\mu\right)
\right|0\right>_+
\end{equation} 
where the subscript \begin{math} + \end{math} denotes path ordering. 
The Wilson conjecture is that the action is proportional to the area 
for sufficiently large loops; i.e. 
\begin{equation}
W=(\tau /c)\Sigma .
\end{equation}
In order to see the physical significance of the proportionality 
constant \begin{math} \tau  \end{math}, consider a rectangular 
space-time Wilson loop as shown in Fig.\ 1. The particle moves forward 
in time \begin{math} T \end{math} as a quark, space-like 
displacement \begin{math} R \end{math} as a string, backward 
in time \begin{math} -T \end{math} as an anti-quark, 
and space-like displacement \begin{math} -R \end{math} as a string. 
The area of the loop is 
\begin{equation}
\Sigma =-cRT,
\end{equation} 
where the negative sign comes from the factor 
\begin{math} (-1) \end{math} in the Lorentz metric for time-like 
intervals. The Wilson loop action for this case is then given by
\begin{equation}
W=-\tau RT.
\end{equation}

\begin{figure}[htbp]
\begin{center}
\mbox{\epsfig{file=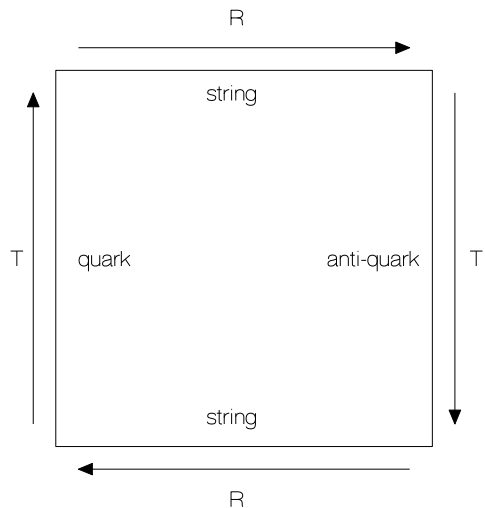,height=80mm}}
\caption{In the above rectangular Wilson loop in space-time, the quark 
moves forward in time while the anti-quark moves backward in 
time. On the space-like top and bottom sides of the loop, the 
superluminal quark path forms a string. The oriented area enclosed by the 
rectangular loop is $\Sigma =-cRT$ and the action $W=(\tau /c) \Sigma $.}
\label{virfig1}
\end{center}
\end{figure}

For an observer of the loop, the physical situation is as follows: 
There exists a quark and an anti-quark at fixed points in space  
connected by a string of length \begin{math} R \end{math}. 
The string energy \begin{math} E_{string} \end{math} 
for this situation is contained in the action \begin{math} W \end{math} 
via the quantum mechanical rule for the time translation amplitude,  
\begin{equation}
\exp(iW/\hbar )=\exp(-iE_{string}T/\hbar ).
\end{equation}
The energy of the string of length \begin{math} R \end{math} is 
thereby 
\begin{equation}
E_{string}=\tau R.
\end{equation}
The parameter \begin{math} \tau \end{math} in the Wilson loop Eq.(2) 
is merely the string tension.

If the Wilson conjecture holds true, then a meson which consists 
of a quark (at position \begin{math} {\bf r}_1  \end{math}) 
and an anti-quark (at position \begin{math} {\bf r}_2 \end{math}) 
exhibits confinement from a ``string potential energy''  
\begin{equation}
U_{meson}({\bf r}_1,{\bf r}_2)=\tau |{\bf r}_1-{\bf r}_2|.
\end{equation}
More generally, if the action of Wilson loops scales as the 
area, then a potential energy for \begin{math} n \end{math} 
constituent quarks obeys the scaling law  
\begin{equation}
U(\xi {\bf r}_1,...,\xi {\bf r}_n)=\xi^h U({\bf r}_1,...,{\bf r}_n), 
\end{equation}
where 
\begin{equation}
h=1\ \ \ {\rm (Wilson\ Strings)}.
\end{equation}

The purpose of this work is to examine the consequences of homogeneous 
scaling potentials for the partitioning of the hadron rest energy 
\begin{math} Mc^2  \end{math} into quark energy and glue energy. 
If the hadron is made up of light quarks, then it turns out   
that Wilson scaling implies that the rest energy of the hadron is in 
equal parts quark energy and glue energy. This equipartitioning 
of hadron rest energy is in experimental agreement with strong 
interaction data, as will be discussed in Sec.\ IV. If the hadron 
contains heavy quarks, then Wilson scaling predicts deviations from 
the equipartition of energy rule. 

In Sec.\ II, we consider the virial for an arbitrary number of
quarks and derive a relativistic virial theorem which relates the 
total mean $4$-momentum of the quarks to that of the glue and an 
spatial integral over the trace of the quark energy-momentum tensor. 
In Sec.\ III, we show that for a hadron made of light quarks 
(i.e., in the limit of vanishing quark masses) obeying Wilson 
scaling, the mean energy carried by the quarks is equal to that
carried by the glue. In Sec.\ IV, we review briefly the well known 
data on the ``parton'' distributions in the best studied
case of the proton confirming our theoretical results obtained in 
Sec.\ III. We close the paper with some concluding remarks in 
Sec.\ V.

\section{Relativistic Virial Theorem}

If the positions and momenta of the quarks are denoted by 
\begin{math} ({\bf r}_1,...{\bf r}_n,{\bf p}_1,...{\bf p}_n) \end{math} 
and if, in the center of mass frame, the positions and momenta of the 
quarks are bounded, then the virial 
\begin{equation}
{\cal V}=(1/2)\sum_{j=1}^n 
({\bf r}_j{\bf \cdot p}_j+{\bf p}_j{\bf \cdot r}_j)
\end{equation}
is also bounded. The internal (center of mass frame) 
wave function \begin{math} \Psi \end{math} of a hadron having 
mass \begin{math} M \end{math} obeys 
\begin{equation}
{\cal H}\Psi =Mc^2\Psi ; 
\end{equation} 
Hence,  
\begin{equation}
\left<\Psi ,\dot{\cal V} \Psi \right>=
(i/\hbar )\left<\Psi, \big[ {\cal H},{\cal V}\big]\Psi \right>=0.
\end{equation}
The rate of change (in time) of the virial Eq.(10) obeys 
\begin{equation}
\left<\Psi ,\dot{\cal V} \Psi \right>
=\left<\Psi ,\sum_{j=1}^n ({\bf v}_j{\bf \cdot p}_j
+{\bf r}_j{\bf \cdot f}_j)\Psi \right>=0, 
\end{equation}
where the velocity 
\begin{equation}
{\bf v}_j=(i/\hbar )\big[ {\cal H},{\bf r}_j\big]=
(\partial H_j/\partial {\bf p}_j)=c{\bf \alpha }_j
\end{equation}
is calculated from the Dirac Hamiltonian 
\begin{math} H_j \end{math} of the 
\begin{math} j^{th}  \end{math} quark. If the quark 
mass is \begin{math} m_j  \end{math}, then 
\begin{equation}
H_j=c{\bf \alpha }_j{\bf \cdot p}_j+m_jc^2\beta_j .
\end{equation}
The force on the \begin{math} j^{th}  \end{math} quark   
\begin{equation}
{\bf f}_j=\dot{\bf p}_j=(i/\hbar )\big[ {\cal H},{\bf p}_j\big]
\end{equation}
is calculated from a potential of the binding glue  
\begin{equation}
{\bf f}_j=-{\bf grad}_j U.
\end{equation}
The potential obeys the Wilson scaling Eqs.(8) and (9).

From the nature of the Dirac Hamiltonian Eq.(15), it follows that 
\begin{equation}
\sum_{j=1}^n{\bf p}_j{\bf \cdot }{\bf v}_j+\sum_{j=1}^n m_j
(\partial H_j/\partial m_j)=\sum_{j=1}^n H_j,
\end{equation}
while the Euler equation corresponding to Eq.(8) is well known to be  
\begin{equation}
\sum_{j=1}^n {\bf r}_j{\bf \cdot f}_j=
-\sum_{j=1}^n {\bf r}_j{\bf \cdot grad}_jU=-hU.
\end{equation}
The mean energy of the quarks, 
\begin{equation}
\bar{E}_{quarks}=\left<\Psi,\sum_{j=1}^n H_j \Psi \right>,
\end{equation}
and the mean energy of the glue, 
\begin{equation}
\bar{E}_{glue}=\left<\Psi, U \Psi \right>,
\end{equation}
are related via Eqs.(13) and (18-21) as in the following   
\medskip 
\par \noindent 
{\bf Virial Theorem:} 
\begin{equation}
\bar{E}_{quarks}=h\bar{E}_{glue}+
\left<\Psi ,\sum_{j=1}^n m_j(\partial H_j/\partial m_j)\Psi \right>.
\end{equation}

The second term on the right hand side of Eq.(22) is related to the 
trace of the quark contribution \begin{math} T^{\mu \nu} \end{math} 
to the mean local hadron stress-energy 
tensor; i.e.  
\begin{equation}
-\int_{hadron} T^\mu_{\ \mu }({\bf r})d^3{\bf r}=
\left<\Psi ,\sum_{j=1}^n m_j(\partial H_j/\partial m_j)\Psi \right>.
\end{equation}
The trace of the stress-energy tensor is simply expressed in 
terms of the local quark energy density 
\begin{math} \varepsilon \end{math} 
and the local quark pressure \begin{math} P \end{math}; 
i.e. 
\begin{equation}
-T^\mu _{\ \mu} ({\bf r})= \varepsilon ({\bf r})-3P({\bf r}). 
\end{equation}
Thus, the virial theorem follows from Eqs.(22-24) to read as    
\begin{equation}
\bar{E}_{quarks}=h\bar{E}_{glue}+\int_{hadron} 
\big\{\varepsilon ({\bf r})-3P({\bf r})\big\}d^3 {\bf r}.
\end{equation}

Finally, let us here relax the restriction of working in the hadron 
center of mass frame. In terms of the total four momentum of the 
hadron 
(\begin{math} P^\mu=\bar{P}^\mu _{quarks}+ \bar{P}^\mu _{glue}\end{math}),
\begin{equation}
\bar{P}^\mu _{quarks}=h \bar{P}^\mu _{glue}-
\left({1\over c}\right)
\int_{hadron} T^\lambda _{\ \lambda}(x) d^3\sigma ^\mu (x), 
\end{equation}
where the four vector ``3-volume components'' are defined via  
\begin{equation} 
d^3\sigma_\mu (x)=\left({1\over 3!}\right)\epsilon_{\mu \nu \lambda \beta
}
dx^\nu \wedge dx^\lambda \wedge dx^\beta .  
\end{equation}
The above manifestly Lorentz invariant form of the virial theorem 
Eq.(26) is the central result of this section.

\section{Light Quark Equipartition of Hadron Energies}

For hadrons made up of light quarks, one expects the 
``ultra-relativistic'' quark gas result 
\begin{math} \varepsilon \approx 3P  \end{math}. The precise 
theorem is that in the limit of vanishing quark masses  
\begin{equation}
-\lim_{m\to 0}T^\mu_{\ \mu} =
\lim_{m\to 0}\left(\epsilon-3P\right)=0.
\end{equation}
Eqs.(9), (25) and (28) imply the equipartition of the hadron 
rest energy 
\begin{math} Mc^2=\bar{E}_{quark}+\bar{E}_{glue} \end{math}
in Wilson loop string models; i.e.  
\begin{equation}
\bar{E}_{quark}=\bar{E}_{glue}\ \ \ 
{\rm (light\ quarks\ on\ strings)}.
\end{equation}

To see simply how Eq.(29) arises, let us consider a simple model 
of mesons made up of a quark and an anti-quark and let us employ 
the string potential of Eq.(7); 
\begin{math} U_{meson}(R)=\tau R \end{math} where 
\begin{math} R \end{math} is the length of the string. 
The ultra-relativistic quark kinetic energy corresponding to 
a meson angular momentum \begin{math} J  \end{math} is given 
by \begin{math} K_{meson}(R)=2c(J/R) \end{math}. 
The mass of the meson may be estimated by the minimum energy 
principle \cite{4,5,6}  
\begin{equation}
M_Jc^2=\inf_{0<R<\infty }\left\{K_{meson}(R)+U_{meson}(R)\right\}.
\end{equation}
The physical string length \begin{math} R_J \end{math} then obeys 
\begin{equation}
R_J^2=(2cJ/\tau ),
\end{equation}
while the mass squared determines a linear angular momentum trajectory 
\begin{equation}
M_J^2=(8\tau J/c^3).
\end{equation}
For the above (perhaps overly) simple model, note that  
\begin{math} \bar{E}_{quark}=K_{meson}(R_J)
=U_{meson}(R_J)=\bar{E}_{glue}\end{math}. 
Thus, Eq.(29) holds true. 

However, the equipartition of energy in Eq.(29) is {\em not} a 
trivial physics result. Different models yield different partitions of 
the energy. For some ``bag'' models,\cite{7} 
\begin{math} U_{meson}(R;{\rm bag})\approx (4\pi /3)BR^3 \end{math}, 
or more generally 
\begin{equation}
h\approx 3 \ \ \ {\rm (bag\ models)}.
\end{equation}
Applying the virial theorem for this case yields 
\begin{equation}
\bar{E}_{quark}\approx 3\bar{E}_{glue}\ \ \ 
{\rm (light\ quarks\ in\ bags)}.
\end{equation}
On the basis of Eqs.(29) and (34) and the discussion in 
Sec.\ IV, it appears that string models are experimentally 
favored over bag models. 

\section{``Parton'' Distribution Data}

Our results from  Secs. II and III are that for light quark
bound states such as the proton and the pion, we expect the
$50 \div 50$ rule for the momenta carried by the quarks and the glue.
Here we shall briefly discuss an experimental verification of
this fact for the proton in the context of the parton model. 
As well known (see for example \cite{8}\cite{9}), one defines a quark
density $q(x)$, and antiquark density $\bar{q}(x)$ and a gluon
density $G(x)$, where $x$ is the fraction of the proton momentum.
What has been found experimentally is that
\begin{equation}
\sum_q \int_o^1 (dx)[q(x) + \bar{q}(x)] \approx 0.5 .
\end{equation}

To quote ref.[9], ``[the above] is an experimental result. It indicates
that the quarks carry about $50\%$ of the proton's momentum. The rest
is attributed to gluon constituents.''

There are similar albeit less accurate data for the pion structure
function as well showing the equipartition of the pion momenta
into quarks and glue.

We expect sizeable deviations from the equipartition for large
quark masses. Present data for heavy quarks are not sufficiently
accurate to  allow for a quantitative test.

\section{Conclusions}

	In the preceding, we have derived a relativistic virial
theorem valid for a generic hadron. For homogeneously scaling 
potentials, it connects the mean energy carried by the quarks, 
the glue and the trace of quark energy-momentum tensor integrated 
over the hadronic volume. For ``ordinary'' hadrons such as the 
pion and the nucleon, made of light quarks, it leads to the pleasing 
result that for the string model (which satisfies the Wilson area 
conjecture) the mean quark energy equals the mean energy carried 
by the glue. There is ample experimental evidence via the quark 
and gluon densities verifying this $50 \div 50$ rule for the case of the 
proton. (A less precise but positive confirmation is also found 
experimentally for the pion). The ``bag models'' generally fail in 
this respect.

\end{document}